\documentclass[sigconf, 10pt, screen, nonacm]{acmart}
\usepackage[subtle]{savetrees}

\usepackage{epsfig}
\usepackage{balance}
\usepackage{graphicx}
\usepackage{subcaption}
\usepackage{tabularx}

\usepackage{xspace}
\usepackage{listings}
\usepackage{verbatim}
\usepackage{booktabs}
\usepackage{colortbl}

\usepackage{nicefrac}
\usepackage{siunitx}

\usepackage{amsmath}
\usepackage{amsfonts}
\usepackage{xfrac}

\usepackage{color}
\definecolor{darkred}{rgb}{0.7,0,0}
\definecolor{darkgreen}{rgb}{0,0.5,0}
\hypersetup{colorlinks=true,
  linkcolor=darkred,
citecolor=darkgreen}

\usepackage{fancyvrb}
\usepackage{relsize}

\usepackage{algorithm,algorithmicx}
\usepackage[noend]{algpseudocode}
\usepackage[normalem]{ulem}

\usepackage{wrapfig}
\usepackage{blindtext}
\usepackage{makecell}
\usepackage{lipsum}

\usepackage{enumitem}
\usepackage{cleveref}
\usepackage{multirow}

\newcommand{\eg}{{e.g.,} }

\newcommand{\kz}[1]{{\textcolor{purple}{KZ: {#1}}}}

\newcommand{\rev}[1] {#1}

\newcommand{\cut}[1]{}

\newcommand{\Para}[1]{\vspace{4pt}\noindent{\bf #1}\hspace{6pt}}

\newcommand{\ours}{\texttt{Polyphony}\xspace}
\newcommand{\polym}{\texttt{pol/m3}\xspace}
\newcommand{\polyp}{\texttt{pol/pmn}\xspace}
\newcommand{\polys}{\texttt{SelfTune}\xspace}
\newcommand{\mthree}{{m3}\xspace}
\newcommand{\pmn}{{Parsimon}\xspace}

\newcommand{\nsthree}{{ns-3}\xspace}
\newcommand{\cloudlab}{{CloudLab}\xspace}

\algrenewcommand\algorithmicrequire{\textbf{Precondition:}}

\def\compactify{\itemsep=0pt \topsep=0pt \partopsep=0pt \parsep=0pt}
\let\latexusecounter=\usecounter

\begin{document}

\title{\LARGE Prediction-Guided Control in Data Center Networks}

\author{\large Kevin Zhao$^{1}$, Chenning Li$^{2}$, Anton A. Zabreyko$^{2}$,
  Arash Nasr-Esfahany$^{2}$, Anna Goncharenko$^{1}$,\\ David Dai$^{1}$, Sidharth
  Lakshmanan$^{1}$, Claire Li$^{1}$, Mohammad Alizadeh$^{2}$, Thomas E.
Anderson$^{1}$\\{$^{1}$University of Washington, $^{2}$MIT CSAIL}}

\renewcommand{\shortauthors}{Zhao, et al.}

\begin{abstract}
  In this paper, we design, implement, and evaluate \ours, a system to give network operators
a new way to control and reduce the frequency of poor tail latency events in multi-class 
data center networks,
on the time scale of minutes.  \ours is designed to be complementary to other adaptive mechanisms like 
congestion control and traffic engineering, but targets different aspects of network operation that have 
previously been considered static.  By contrast to \ours, prior model-free optimization methods 
work best when there are only a few relevant degrees of freedom and where workloads and measurements are stable,
assumptions not present in modern data center networks. 

\ours develops novel methods for measuring, predicting, and controlling network quality
of service metrics for a dynamically changing workload.  First, we monitor and aggregate workloads
on a network-wide basis; we use the result as input to an approximate counterfactual prediction engine that
estimates the effect of potential network configuration changes on network quality of service; we apply
the best candidate and repeat in a closed-loop manner aimed at rapidly and stably converging to a
configuration that meets operator goals.
%
Using CloudLab on a simple topology, we observe that \ours converges to tight SLOs within ten minutes, and
re-stabilizes after large workload shifts within fifteen minutes, while the prior state of the art fails to adapt.

\end{abstract}

\maketitle

\pagestyle{plain}
\section{Introduction}

Improving network quality of service --- the likelihood that network transfers complete in a timely fashion ---
is a longstanding goal of networking research~\cite{intserv,shenkerqos,clarkrealtime,diffserv,shenkerpricing}.
Data center applications in particular are often very latency sensitive~\cite{dcqcn,azurerdma,azuresmartnic,snap},
with inherent dynamic variability~\cite{jupiterevolving}, a predominance of short flows~\cite{bfc}, and
frequent incast and outcast traffic patterns~\cite{jupiter}. Link oversubscription to reduce costs, as well as
link and switch failures add to the challenges.
A number of techniques have been proposed, including resource reservations~\cite{intserv}, priority scheduling~\cite{diffserv},
pricing signals to discourage usage during periods of contention~\cite{shenkerpricing}, receiver-based management of incast communication~\cite{ndp,homa}, switch behavior to bypass queues~\cite{csfq,hull,qjump,cal-queues,spfifo,bfc}, and central coordination of when hosts are allowed to transmit~\cite{b4,fastpass}.

In practice, most production data center networks combine three ideas to improve network quality of service, albeit
with mixed results. First, aggressive congestion control algorithms are designed to react quickly (within a few round trips)
and forcefully to observed congestion along a path in an attempt to keep queues small~\cite{dctcp,hpcc,swift,hull}.
Second, on longer time scales, traffic engineering (such as weighted ECMP and Optical
Circuit Switching (OCS)) attempts to balance traffic volumes over available paths despite non-uniform topologies and inherent
temporal and spatial variability in traffic demand~\cite{jupiterevolving}.
Third, as a final backstop, mission critical or latency sensitive traffic is assigned to separate traffic classes to isolate
it from other competing traffic, typically using scheduling weights rather than priorities to avoid starvation for low priority traffic.
Even with these mechanisms in place, well-managed data center networks still
experience significant tail slowdowns for RDMA and RPC operations caused by network congestion~\cite{azurerdma,googlerpc}.

In this paper, we design, implement, and evaluate \ours{}, a system to give network operators
a new way to control and reduce the frequency of poor tail latency events in multi-class data center networks.
\ours{} is designed to be orthogonal and complementary to other workload-adaptive mechanisms like
congestion control and traffic engineering. We assume the network operator sets some traffic class specific performance goal,
such as keeping the 99th percentile flow completion time (FCT) slowdown --- defined as FCT normalized to its
lowest value in an unloaded network --- below a threshold.

What additional mechanisms do operators have to effectively control the FCT?
Modern data center networks expose a broad control surface, ranging from various
end-host congestion control parameters to switch-level queueing, scheduling, and buffer
allocation policies. Today, these parameters are typically configured once, or adjusted only in response to
emergencies.
However, we show that the effects on tail latency can be profound: small changes in end host
or switch behavior can reshape latency distributions to achieve, or fail to achieve, operator goals.
Further, the appropriate changes to make are highly workload dependent and therefore time varying -- burstiness, correlated behavior,
and even the behavior of other traffic classes can affect user-observable outcomes, in ways that
are hard to define from first principles.

Our approach is to develop a new {\em prediction-guided controller} for data center networks
that rapidly converges, in response to workload or topology changes, over a period of minutes.
Unlike purely reactive congestion control that collects evidence
on a single path at a time, \ours{} samples and aggregates network workload data on a network-wide basis
to reduce the likelihood of future congestion events {\em before} they happen.
Unlike traffic engineering, \ours{} samples much richer workload information than just traffic volumes.

We leverage recent advances in fast approximate models for estimating network performance.
These models predict tail latency for a given network workload, topology, and configuration~\cite{mimicnet,deepqueuenet,zhao2023scalable,li2024m3,gao2023dons}.
They can produce approximate answers for how network performance may change as a result of some control action,
many orders of magnitude faster than detailed packet-level simulation frameworks like ns-3~\cite{ns-3}.
Specifically, we use two open source models, Parsimon~\cite{zhao2023scalable} and m3~\cite{li2024m3}.
We extend them to support class-based queueing and to work on live network traffic, specifically CloudLab~\cite{duplyakin2019design}.
Because m3 uses machine learning, we train it using CloudLab
measurements as ground truth; the original work trained against ns-3. CloudLab introduces
various sources of variability not present in ns-3, such as host jitter that measurably affects how the network is used.
As a result, our CloudLab-trained m3 is substantially more accurate than both
the original m3 and Parsimon. This allows us to evaluate robustness to
prediction error.
We find that a slower and less accurate predictor slows convergence.
We show this sensitivity in our experiments, using m3 as the higher-fidelity
predictor and Parsimon as a coarser baseline.

The general approach of prediction-guided network control poses several problems which this work addresses.
First, network prediction models have significant error, and we show that this means
they cannot be used directly to recommend configurations (or if we do, then we end up stuck at a non-optimal operating point).
Rather, as we explore the configuration space, we train a Gaussian model to correct for
observed model error between the predicted and live system. Since this corrected model is most accurate in the neighborhood
of previous trials, we bias the search to favor exploration within a trust region around the best known
configuration. Finally, we use a denoiser to smooth the results across different trials to reduce sensitivity
to measurement variance.  We show through an ablation study that each of these steps is necessary to efficiently
reach a good operating point.

As a proof of concept, we evaluate \ours{} using each of the two prediction models on CloudLab for a small network, and the more accurate m3 model
on ns-3 to study behavior on larger-scale networks, varying nine network configuration parameters.
Relative to SelfTune, a state of the art model-free optimizer, given tight class-specific tail latency goals and starting
from a non-optimal configuration, \ours{} using m3 converges on CloudLab to a configuration that meets the objective within
ten minutes of real time, while SelfTune does not make much progress even within the sixty-minute trial period.
%
When we introduce large periodic workload changes, SelfTune shows no clear signs of adaptation, while
\ours re-stabilizes to SLOs within fifteen minutes.
{\ours}'s source code will be publicly available.

\section{Background and Motivation} \label{s:background}

\begin{figure*}[t]
  \centering
  \begin{subfigure}[t]{0.24\linewidth}
    \centering
    \includegraphics[width=\textwidth]{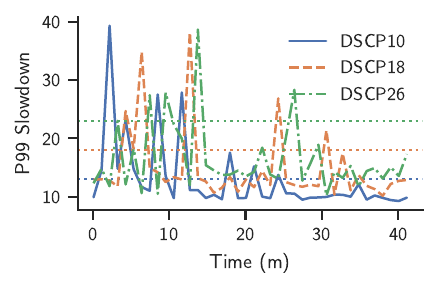}
    \vspace{-5mm}
    \caption{\small P99 FCT Slowdowns}
    \label{fig:ego-slowdown}
  \end{subfigure}
  \hfill
  \begin{subfigure}[t]{0.24\linewidth}
    \centering
    \includegraphics[width=\textwidth]{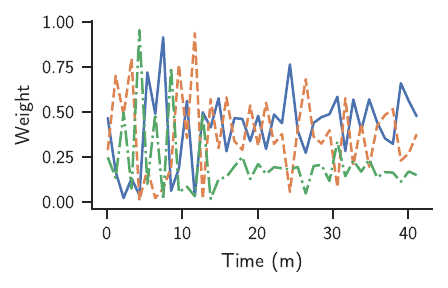}
    \vspace{-5mm}
    \caption{\small Switch weights}
    \label{fig:ego-weight}
  \end{subfigure}
  \hfill
  \begin{subfigure}[t]{0.24\linewidth}
    \centering
    \includegraphics[width=\textwidth]{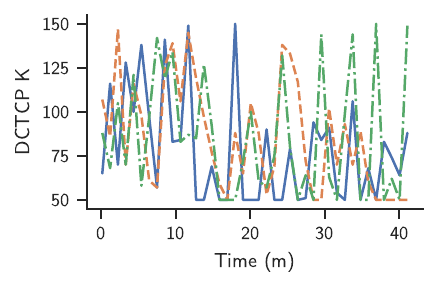}
    \vspace{-5mm}
    \caption{\small DCTCP K}
    \label{fig:ego-k}
  \end{subfigure}
  \hfill
  \begin{subfigure}[t]{0.24\linewidth}
    \centering
    \includegraphics[width=\textwidth]{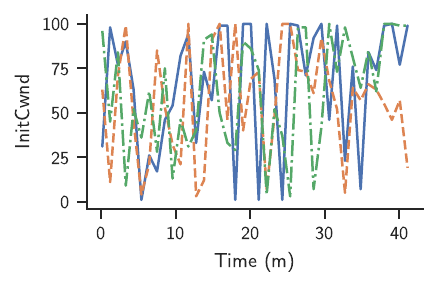}
    \vspace{-5mm}
    \caption{\small Init CWND}
    \label{fig:ego-win}
  \end{subfigure}
  \vspace{-1mm}
  \caption{\small
    Model-free Bayesian optimization leads to poor behavior during convergence.
    \Cref{fig:ego-slowdown} shows the P99 FCT slowdown of three traffic classes over time.
    Horizontal lines indicate SLO thresholds.
    Both configurations and tail FCT slowdowns oscillate as the
    optimizer explores the parameter space.
  }
  \vspace{-2mm}
  \label{fig:ego}
\end{figure*}

\begin{figure}[t]
  \centering
  \begin{subfigure}[t]{0.48\linewidth}
    \centering
    \includegraphics[width=\linewidth]{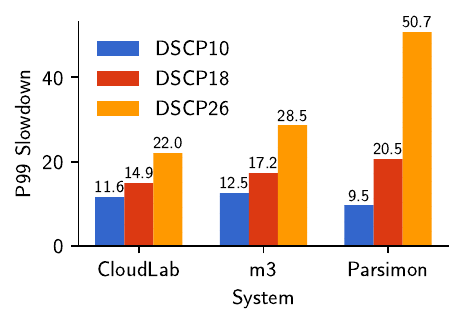}
    \phantomsubcaption\label{fig:model-error}
  \end{subfigure}
  \hfill
  \begin{subfigure}[t]{0.49\linewidth}
    \centering
    \includegraphics[width=\linewidth]{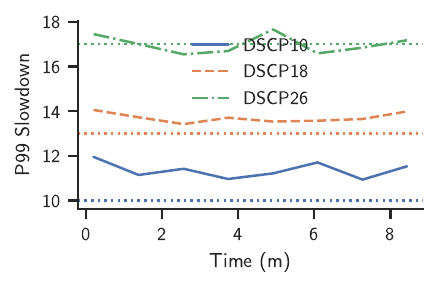}
    \phantomsubcaption\label{fig:open-loop-slowdown}
  \end{subfigure}
  \vspace{-6mm}
  \caption{\small
    Model error and its consequences for three classes with varying SLOs. \mthree is trained on CloudLab
    while Parsimon uses packet-level simulation.
    (a) Even \mthree mispredicts CloudLab's P99 FCT
    slowdown by ~18\% on average.
    (b) When the best configuration predicted by \mthree is applied to the network,
    all three traffic classes violate their SLOs, even though they can all be met simultaneously.
    The configuration is kept the same across
    samples; variance is due to noise.
  }
  \label{fig:model-vs-openloop}
  \vspace{-4mm}
\end{figure}

%
The setting for our work is a network where traffic is
divided into multiple service classes, each with its own tail latency objective.
For example, an operator might require that 99\% of flows in
one class must have a flow completion time (FCT) slowdown no greater than 10$\times$, while another class
may be able to tolerate a slowdown up to 20$\times$.
Following Mogul and Wilkes~\cite{mogul-nines}, we call this target a service
level objective (SLO), and the measured performance metric (P99 FCT slowdown) a
service level indicator (SLI).
The bound (e.g., 10 or 20$\times$) is the SLO threshold.

Our approach is agnostic to the unit of measurement, but to be concrete
we assume it is a remote procedure call (RPC), remote memory operation (RDMA), or independent
data transfer. Latency is measured as the time to complete the transfer, including
the transmission, propagation, and queueing delay, from when the first packet is
available to be sent until the last packet arrives at the destination.
We include in the latency any queueing at the end host
needed for traffic shaping or congestion control.
Following industry practice, we modify the host operating system to
sample tail latencies as input to our control system.

We define FCT slowdown as the flow latency divided by the minimum
latency on an unloaded network. For example, in a network with a round trip
propagation delay of 10\,$\mu$sec and 10\,Gbps links, the minimum latency
for a 12.5KB transfer would be 20$\mu$sec. \ours{} supports performance targets to be defined
separately for different message sizes, e.g., to ensure that medium-sized messages aren't starved.
In this paper, we focus on aggregate tail performance.

We assume all network hosts and switches are configured identically, but in a class-aware manner.
Each traffic class is assigned its own dedicated queue at each network link; each class is given a scheduling weight that controls
how frequently packets from that class are scheduled when other traffic is present.
All hosts and switches implement the same congestion control algorithm, but the parameters controlling the behavior
of the algorithm, such as the initial window size or its overall aggressiveness, are traffic class specific.
The prediction models we use generalize across many commonly used congestion control
algorithms~\cite{zhao2023scalable,li2024m3}, but for our proof of concept we narrow our focus to DCTCP~\cite{dctcp}.
It contains an initial congestion window and a threshold parameter, $K$, for switch queues to trigger end hosts to reduce their sending rate;
these parameters are allowed to vary by traffic class.

Whether a specific traffic class meets its objective depends on the specific workload and parameters of that class,
but also that of other classes.  Class-based switch scheduling allocates a worst case minimum share of the line rate to each
class in proportion to its weight, but scheduling is also work conserving.  If a class does not have traffic present (e.g., because it is bursty)
the unused capacity is proportionately split among the other traffic classes with traffic present.
As a result, the effect of individual changes may not be monotonic.  For example, the DCTCP $K$ value controls
the target amount of queueing at congested switches.  Reducing this value can improve tail latency for flows that fit within
the initial congestion window, but it can also hurt tail latency by reducing the ability of the traffic class to take advantage
of the idle periods of other classes.

\subsection{Existing methods} \label{s:existing}

\rev{

  Given that the network exposes parameters for control, our goal is not a ``best
  configuration'' but continuous regulation: adjusting settings in closed loop to
  track SLOs as workloads evolve.
  When operating conditions change---for example a shift in traffic patterns or
  workload intensity---the controller should adapt while minimizing SLO violations.
  We therefore evaluate approaches by: (i) convergence time (time to reach
  per-class SLO compliance), (ii) regret (cumulative deviation from compliance
  during adaptation), and (iii) fairness (whether deviations are borne evenly
  across classes).
  In our experiments, we treat the SLOs for each traffic classes
  as equally important, but \ours{} can also prioritize certain classes with weighted objectives.

  \Para{Model-free autotuners.}
  A prominent line of prior work adjusts parameters using black-box autotuning.
  These methods are model-free: they iteratively deploy settings and observe
  outcomes on the live system.
  For stability, they typically operate on long time horizons, with each
  iteration lasting up to hours or days.
  A common example is Bayesian optimization~\cite{bayesopt}, which selects configurations to test
  by balancing exploration (trying uncertain settings) and exploitation (choosing
  promising ones).
  While model-free methods can eventually find good configurations, they must
  search by running real experiments on the target system, risking SLO violations (regret)
  in the process.
  \Cref{fig:ego} shows standard Bayesian optimization tuning a small CloudLab network
  with three traffic classes and nine parameters (per-class switch weight, DCTCP
  marking threshold, and initial congestion window).
  Although the optimization eventually settles after about thirty minutes, it
  first explores widely, causing users of the system to experience significant SLO violations.
  We would expect these oscillations to be more pronounced with more traffic classes and additional per-class parameters.

  An emerging class of model-free autotuners uses reinforcement learning
  (RL).
  RL tuners like SelfTune~\cite{karthikeyan2023selftune} and
  OPPerTune~\cite{somashekar2024oppertune} update parameters by acting in
  discrete rounds: in each round, the tuner picks a configuration, the system
  runs for some time---typically ranging from an hour to days---and a scalar
  reward is fed back.
  These approaches rely on round-level measurements being information-rich and
  averaging out noise.
  In our setting with short timescales and noisy tail latency measurements, these
  conditions do not hold.
  As a result, reward signals can be noisy and gradients weak, causing updates to
  shrink and learning to stall.
  We apply SelfTune to our setting in the evaluation (\Cref{s:eval}).
  We find that, whether by Bayesian optimization or reinforcement learning,
  model-free autotuning struggles to track tail latency SLOs on short timescales
  in dynamic, noisy environments.

  \Para{Emerging fast models.}
  Meanwhile, recent projects like Parsimon and m3
  have developed fast, approximate simulation methods for predicting tail latency
  in data center networks.
  One idea would be to use their predictions to drive configurations directly in
  open loop.
  Unfortunately, as \Cref{fig:model-error} shows, these models still have significant
  error.
  Even m3, which can be trained directly on the target platform, exhibits in this
  scenario an average of about 18\% error in predicting 99th percentile FCT slowdown
  across traffic classes.
  \Cref{fig:open-loop-slowdown} shows what happens when we use the model to find
  its predicted best configuration, and then apply it directly.
  Even though the SLOs are achievable, the deployed configuration misses all
  SLOs.
}

\vspace{2mm}

Model-free search learns from live measurements but can incur SLO violations due
to wide exploration; model-based optimization can avoid unsafe exploration but
inherits the model's errors.
Neither alone provides fast, low-regret adaptation for tail latency SLOs.
\ours combines aspects of both approaches by coupling a fast model with
online corrective feedback.
The next section outlines \ours's closed-loop architecture and introduces how
it blends simulation and live measurement to achieve SLO compliance.

\section{\ours Overview} \label{s:overview}

Given a live network that serves multiple traffic classes with per-class tail latency SLOs, our goal is to continuously adjust a set of knobs---like switch weights or congestion control parameters---so all SLOs are met and tail latency is minimized.
As a secondary goal, allocations should be \emph{fair} in the sense that no class is persistently disadvantaged relative to its SLO.
We frame this as a regulation problem: \ours acts as an online controller that tracks per-class SLO constraints via small, bounded adjustments.

\begin{figure}[t]
  \centering
  \includegraphics[width=0.99\linewidth]{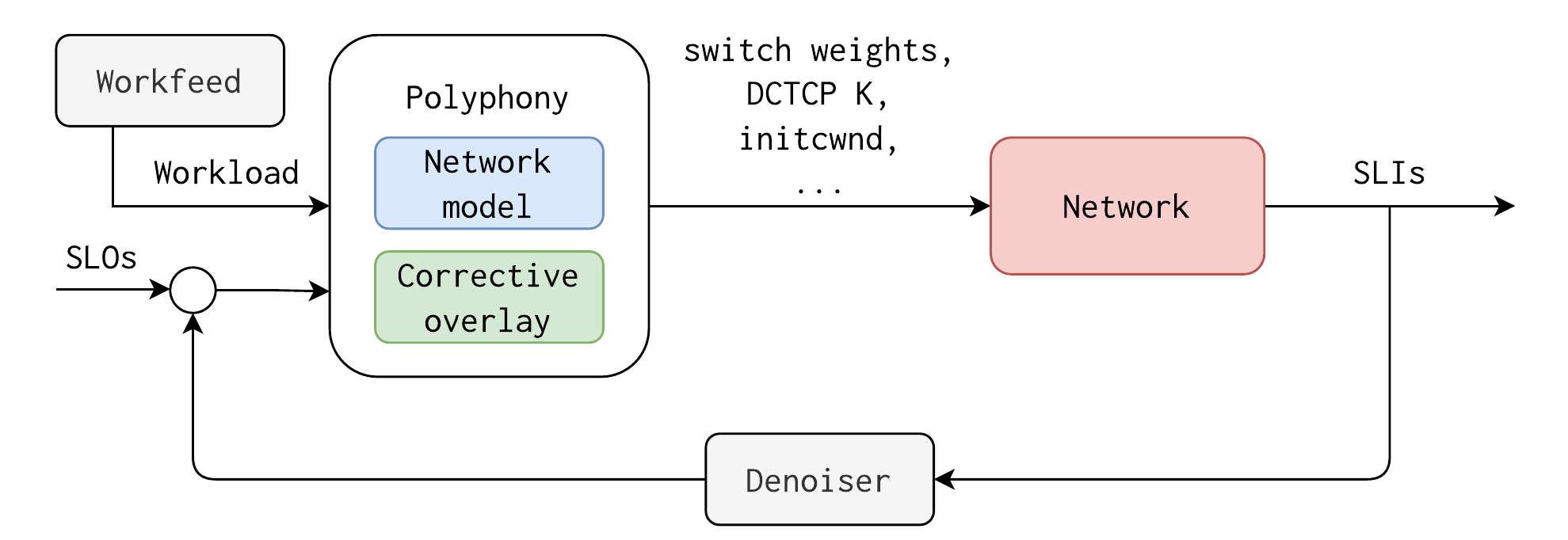}
  \vspace{-2mm}
  \caption{\small
    \ours's closed-loop architecture.
  }
  \vspace{-4mm}
  \label{fig:overview}
\end{figure}

\Cref{fig:overview} depicts \ours's closed-loop architecture.
The core idea is to combine a cheap, approximate performance model with an online-learned corrective overlay.
\ours then uses the overlay to compute the next bounded adjustment within a trust region for a network setting to try next.
By optimizing a predictive model rather than the live system, \ours reduces the number of costly system measurements required to find good configurations, and limits the risk of applying undesirable settings.
%
%
\rev{
  The system takes as input the target SLOs and the current service-level indicators (SLIs), and it outputs the next setting to apply.
  The performance models require workload information as input.
  To measure workloads, \ours bundles a lightweight \texttt{Workfeed} component
  that uses eBPF to capture flow events and reconstruct flow sequences for the models.
}

At each control interval, \ours computes a bounded update to the control knobs.
First, it collects SLIs, including per-class tail slowdowns.
These metrics capture the performance seen by each class and serve as feedback for updating the corrective overlay.
Next, it filters these measurements to reduce the effects of noise and
transient spikes, producing a smoothed estimate of the system performance.
Using this filtered signal, \ours updates its corrective overlay.
This model captures the discrepancy (i.e., residual) between the predictions of the fast, approximate simulator and the live measurements from the real system.
%
%
We use the estimated residual to improve the predictive accuracy of the overall model.
Then, \ours searches for a set of network knobs within a trust region that maximize the expected performance improvement.
This step balances exploration with caution, ensuring that new settings are only tried if the model predicts an improvement with high confidence.
Finally, the selected configuration is sent to the switches and hosts to actuate the intervention.

The following sections describe \ours's methods in detail.
First, we focus on settings with stable workloads in \Cref{s:methods-1}.
Then, we describe adaptation to workload changes in \Cref{s:methods-2}.
\section{Safe Prediction-Guided Optimization} \label{s:methods-1}

\begin{figure*}[t]
  \centering
  \includegraphics[width=0.95\linewidth]{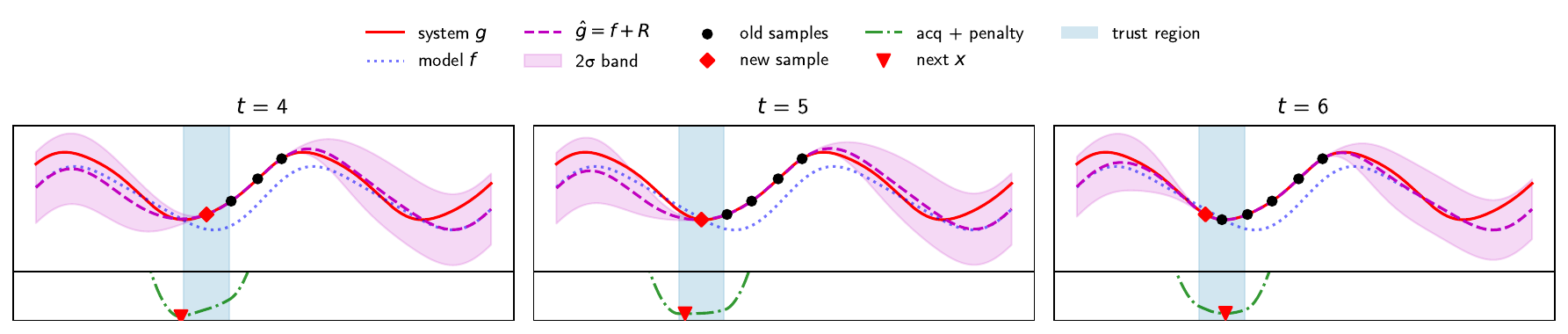}
  \caption{\small
    An illustration of Bayesian optimization with a residual surrogate
    and a trust region.
    Each panel shows one control iteration ($t=4,5,6$).
    The acquisition function (bottom curve) is quadratically penalized outside
    the trust region.
    Over successive steps the surrogate improves near the new sample and the
    trust region recenters, guiding the optimizer toward better configurations
    while avoiding unsafe exploration.
  }
  \label{fig:gp-trust}
  \vspace{-1mm}
\end{figure*}

To start, we view the network as a black box that, for a workload, accepts a vector of parameters $\mathbf{x}$---switch weights, marking thresholds, etc.---and emits a vector of SLIs $\mathbf{y}$, such as per-class tail latencies.
Throughout this section, we assume the workload is fixed and stable, and drop it from notations to avoid clutter.
Alongside the network, we have an approximate model, like \pmn~\cite{zhao2023scalable} or \mthree~\cite{li2024m3}, that takes the same $\mathbf{x}$ and produces \emph{predicted} SLIs for an independent sample of the same workload.
We assume that the model makes errors but is cheap to evaluate, and that the real
system is noisy and risky to probe.
\ours's principal idea is to learn the difference between them and then
optimize the performance using a \emph{corrected} version of the model.
It also optimizes within a safety envelope to avoid straying from known good
configurations.
To make these ideas more precise, we begin by introducing some notation.

\subsection{Definitions and Problem Formulation} \label{s:methods-def}

We first define the parameter space.
\ours considers two kinds of parameters.
The first lie in a hypercube, where each parameter varies independently.
Examples of these are congestion control parameters or initial congestion
windows.
The second lie in a simplex: they are constrained to be non-negative and sum to
one.
These represent allocations or proportions, such as switch weights which must
be distributed among traffic classes.
Polyphony manipulates a vector
$$
\mathbf{x}\;=\;\bigl(\,h^{(1)},\,h^{(2)},\;\dots,\,h^{(M_h)},\;
u^{(1)},\,u^{(2)},\;\dots,\,u^{(M_u)}\bigr),
$$
where $h^{(k)} \in [0,1]^{d_k}$ are hypercube parameters and
$u^{(\ell)} \in \Delta^{d_\ell}$ are simplex parameters.
Combining these domains, the full parameter space is
$$
\mathcal{X}\;=\;\Bigl(\prod_{k}[0,1]^{d_k}\Bigr)
\times\Bigl(\prod_{\ell}\Delta^{d_\ell}\Bigr).
$$
All coordinates are scaled to $[0,1]$ for optimization and inverse-normalized for actuation.

For the particular workload, we think of the system as a function $s(\mathbf{x}): \mathcal{X} \to \mathbb{R}^N$
which maps a vector of parameters to a vector of SLIs.
The model $m(\mathbf{x}): \mathcal{X} \to \mathbb{R}^N$ is defined similarly for the same workload.
Although $s(\mathbf{x})$ and $m(\mathbf{x})$ each return a $N$-dimensional
vector of SLIs, an optimizer needs a total order over configurations to decide
whether one setting is better than another.
\ours therefore collapses the vector into a single scalar cost $J$.
Doing so has two main advantages: 1) it reduces the problem to scalar
optimization with a clear improvement signal, and 2) it allows us to express
SLO compliance and fairness as a single quantity to optimize instead of
exploring a Pareto frontier.

To guide optimization, we need a scalar objective that reflects our goals of
SLO compliance and fairness.
Let $\mathbf{y} = (y_1, \dots, y_N) \in \mathbb{R}^N$ be the observed SLIs
(e.g., tail latencies) for each of the $N$ traffic classes, and let $\mathbf{v}
= (v_1, \dots, v_N) \in \mathbb{R}^N$ be their corresponding SLOs thresholds.
To ensure SLO compliance, we want each class's observed slowdown $y_i$ to be
less than its threshold $v_i$, so we define the per-class ratio $r_i = y_i / v_i$.
A value $r_i \leq 1$ indicates compliance.
Optimizing for SLO compliance then becomes mimizing some aggregate of these
ratios across classes.

We choose the maximum as our aggregate because SLO violations are typically
unacceptable even if they affect only a single class.
Using the maximum of the ratios prioritizes reducing the worst violation.
However, maximum is non-differentiable, which prevents the use of gradient-based optimization methods.
To address this, we use LogSumExp (LSE)~\cite[Sec.~3.1.5]{boyd2004convex} as a smooth approximation of maximum, with sharpness controlled by parameter $c > 0$:
$$
\text{LSE}(r_1, \ldots, r_N)
\;=\;
\frac{1}{c}\,
\ln\!\Bigl(\;\sum_{i=1}^{N} \exp\!\bigl(c\,r_i\bigr)\Bigr),
$$

As $c \to \infty$, the function approaches a sharp maximum, while smaller values
of $c$ yield a smoother behavior.

To encourage fair allocations, we add a fairness penalty which tries to equalize
the $r_i$ ratios.
Let $\bar{r} \;=\; \frac{1}{N}\sum_{i=1}^{N} r_i$ be the mean of the ratios. We
define
$$
\text{fairness}(r_1, \ldots, r_N)
\;=\;
\frac{1}{N}\sum_{i=1}^{N} \bigl|\,r_i - \bar{r}\,\bigr|.
$$
The final objective is then
$$
J(r_1, \ldots, r_N)
\;=\;
\text{LSE}(r_1, \ldots, r_N)
\;+\;
\lambda\;
\text{fairness}(r_1, \ldots, r_N),
$$
where $\lambda \geq 0$ is a trade-off constant.
Lower values of $J$ indicate a smaller worst-case slowdown ratio and/or a
fairer spread of ratios across classes, depending on the weight $\lambda$.

Recall that for a particular workload, the system $s(\mathbf{x})$ and model $m(\mathbf{x})$ each return a
vector of SLIs given a configuration $\mathbf{x}$.
To calculate $r_i$ ratios used for objective ($J$) computation, we can use $s(\mathbf{x})$ or $m(\mathbf{x})$.
We call the former $g(\mathbf{x})$ and the latter $f(\mathbf{x})$.
In words, $g(\mathbf{x})$ is the system's observed performance and $f(\mathbf{x})$ is the model's predicted
performance.
%
In what follows, we refer to $g$ as ``the system'' and $f$ as ``the model.''

\subsection{Modeling Residuals} \label{s:methods-gp}
The controller must minimize a system $g$ that is noisy and risky to
sample while taking advantage of a model $f$ that could have significant modeling
error.
Directly learning $g(\mathbf{x})$ is difficult because it can be
highly nonlinear and discontinuous.  Instead,
\ours\ learns the \emph{residual}

$$
R(\mathbf{x}) \;=\; g(\mathbf{x}) - f(\mathbf{x}),
$$
which is typically smoother and smaller in magnitude.

\Para{Gaussian-process surrogate.}
We treat the residual as a random function drawn from a Gaussian process $R
\sim \mathrm{GP}$.
Given a set of residual observations $\{(\mathbf{x}_j,R_j)\}$,
standard GP inference returns a posterior mean $\mu_R(\mathbf{x})$
and variance $\sigma_R^{2}(\mathbf{x})$.
We use $\hat g(\mathbf{x}) = f(\mathbf{x}) + \mu_R(\mathbf{x})$ as the bias-corrected predictor and $\sigma_{\hat g}^{2}(\mathbf{x}) = \sigma_R^{2}(\mathbf{x})$ as its uncertainty.
%
This uncertainty estimate provides the risk budget that supports the
safe optimization strategy that follows.

\subsection{Bayesian Optimization over $\hat g$} \label{s:methods-bo}

Our surrogate $\hat g$ supplies both predictions and calibrated uncertainty.
This leads naturally to the use of Bayesian optimization to balance exploration
and exploitation.

Bayesian optimization (BO) iteratively chooses the next configuration by
maximizing an \emph{acquisition function} built on top of a fast
\emph{surrogate} of the true objective.  In our case
$$
\text{surrogate} \;=\; \hat g(\mathbf{x}) \quad\text{from \S\ref{s:methods-gp}},
$$
and the acquisition is Expected Improvement (EI).

\Para{Expected Improvement.}
Let $g_{\text{best}}\!=\!\min_{j<t} g(\mathbf{x}_j)$ be the smallest
\emph{observed} objective after $t-1$ iterations.  For a candidate
$\mathbf{x}$ with surrogate mean $\mu_{\hat g}$ and standard deviation
$\sigma_{\hat g}$,
$$
\mathrm{EI}(\mathbf{x})
\;=\;
\bigl(g_{\text{best}}-\mu_{\hat g}\bigr)\,\Phi(z)
+ \sigma_{\hat g}\,\phi(z),\qquad
z=\frac{g_{\text{best}}-\mu_{\hat g}}{\sigma_{\hat g}},
$$
where $\Phi(\cdot)$ and $\phi(\cdot)$ are the standard normal CDF and PDF.
EI is the \emph{expected drop in the objective} if we were to sample at
$\mathbf{x}$.
Anchoring EI to the smallest \emph{measured} value guards against model
over-optimism: a point can only show positive improvement if it is expected
to beat what the system has actually achieved, not merely what the model
predicts.
Because \ours is a minimizer, the implementation minimizes
$A(\mathbf{x})=-\mathrm{EI}(\mathbf{x})$:
$$
\mathbf{x}_{t} \;=\; \arg\min_{\mathbf{x}\in\mathcal{X}}\; A(\mathbf{x}).
$$
EI balances exploitation of low $\mu_{\hat g}$ with exploration in regions of
high $\sigma_{\hat g}$, aiming for sample-efficient improvement over time.

\Para{Handling simplexes.}
Bayesian optimization typically operates in an unconstrained, Euclidean space.
To accommodate simplex-valued parameters—such as traffic class weights that
must be non-negative and sum to one—we optimize in the unconstrained pre-image
and map the result through a softmax transformation.
This allows gradient-based methods to search freely while ensuring that the
final parameter vector lies in the simplex.
The surrogate and acquisition functions are defined over the unconstrained
space but always evaluated on the mapped simplex point.

\subsection{Safe Exploration} \label{s:methods-tr}

While Bayesian optimization efficiently explores the parameter space
$\mathcal{X}$ by balancing exploration and exploitation, it does not
guarantee safety.
Probing $g$ can be risky: unconstrained exploration might lead the optimizer to
query points $\mathbf{x}$ that violate SLOs, even if the surrogate $\hat g$
predicts otherwise.

\Para{Trust region.}
To mitigate this risk, \ours uses a safe exploration strategy that builds on
well-understood methods from the optimization literature \cite{TuRBO}.
%
Instead of optimizing the acquisition over the entire parameter space $\mathcal{X}$,
we restrict the search to a \emph{trust region} centered around the \emph{best
configuration} observed so far.
Let $\mathbf{x}^{(t)}_{\text{best}} = \arg\min_{\mathbf{x}_j, j <
t}g(\textbf{x}_j)$ be the parameters corresponding to the best \emph{observed}
system performance $g_{\text{best}} = g(\mathbf{x}^{(t)}_\text{best})$ after
$t-1$ iterations.
The optimizer at step $t$ focuses its search within a neighborhood of
$\mathbf{x}^{(t)}_{\text{best}}$.

\Para{Distance metric.}
Defining this neighborhood requires a distance metric on the mixed parameter
space $\mathcal{X}$.
Each configuration $\mathbf{x}\in\mathcal{X}$ consists of
1) a hypercube part $\mathbf{h} \in \prod_{k=1}^{M_h}[0,1]^{d_k}$, and
2) $M_u$ simplex blocks
$u^{(1)},\dots,u^{(M_u)} \in \Delta^{d_\ell}$.
We measure separation with
$$
d(\mathbf{x},\mathbf{y})
= \sqrt{\,\underbrace{\lVert \mathbf{h}_\mathbf{x}-\mathbf{h}_\mathbf{y}\rVert_{2}^{2}}_{\text{hypercube}}
\;+\;
\alpha\,
\sum_{\ell=1}^{M_u}
\underbrace{d_A\!\bigl(u^{(\ell)}_\mathbf{x},u^{(\ell)}_\mathbf{y}\bigr)^2}_{\text{simplex}}
},
$$
where $d_A$ is the Aitchison distance commonly used for compositional
data~\cite{aitchison1982statistical} and $\alpha\!\ge\!0$ balances the Euclidean and
compositional parts.
The weight $\alpha$ is tuned so that the simplex and hypercube terms contribute
on the same numerical scale.

\Para{Enforcement.}
Enforcing a hard constraint $d(\mathbf{x},\mathbf{x}_{\text{best}}^{(t)}) \le \epsilon$
during optimization can be complex, especially for simplex parameters, which
we optimize in an unconstrained space.
\ours enforces the trust region by adding a penalty term to the acquisition function:

$$
\text{penalty}(\mathbf{x}\,;\,\mathbf{x}_{\text{best}}^{(t)}, \epsilon)
=
\begin{cases}
0 & d(\mathbf{x},\mathbf{x}_{\text{best}}^{(t)}) \le \epsilon,\\[6pt]
\beta \cdot (\,d(\mathbf{x},\mathbf{x}_{\text{best}}^{(t)})-\epsilon\,)^{2} & \text{otherwise},
\end{cases}
$$
where $\epsilon$ is the current trust-region radius and $\beta>0$ controls the penalty
severity.
The optimizer minimizes
$$
A_{\text{TR}}(x)
= -\,\mathrm{EI}(x) +
\text{penalty}(\mathbf{x}\,;\,\mathbf{x}_{\text{best}}^{(t)}, \epsilon)
$$
thereby searching freely \emph{inside} the ball
$d(\mathbf{x},\mathbf{x}_{\text{best}}^{(t)}) \le \epsilon$ while
rapidly discouraging excursions beyond it.
Because the penalty is quadratic and continuous, gradients remain well-behaved.
In our experiments, we set $\beta=10^4$.

\Cref{fig:gp-trust} illustrates the concepts we have introduced so far.
Next, we describe how \ours adapts to new operating regimes, and how it handles
data to remain both reactive and stable over long time horizons.

\section{Adaptation and Data Quality} \label{s:methods-2}


Thus far, we have described methods that allow \ours to converge to good
solutions so long as $g$ remains stationary.
Real networks, however, can experience step changes: traffic mixes shift, hot
spots appear.
To adapt to new regimes, \ours must 1) detect that the regime has changed, and 2)
forget stale data that would mislead the GP.
This section discusses these mechanisms.
In addition, we describe methods to ensure \ours remains effective over long
timescales.

\subsection{Adapting to changing $g$}

Consider how \ours---as currently described---would respond if the network
suddenly experienced a disruption, such as a sudden influx of traffic.
The value of our objective $g$ would spike, prompting the controller to adjust
parameters to achieve SLO conformance.
However, the trust region would still be centered around the old best point,
tying the optimization to that region and causing it not to make progress.
To prevent this, we first maintain a \emph{rolling} best objective rather than
a static one:
$$
g_{\text{best}}(t)
\;=\;\min_{j=t-W_g}^{t-1} g(\mathbf{x}_j),
$$
where $W_g$ is the size of the rolling window.
Next, we implement a simple scheme for objective spike detection and trust
region reset.
If the current measurement exceeds the rolling best objective by a factor of
$\phi$ for $k$ consecutive iterations,
$
g(\mathbf{x}_t) \;>\; \phi \, g_{\text{best}}(t),
$
we treat the event as a regime shift, and we 1) reset the $g_\text{best}$ window
and 2) designate the current configuration as $\mathbf{x}_\text{best}^{(t)}$.

\subsection{Processing data samples}

\Para{Rolling data.}
As \ours runs over time, the number of samples it collects can grow without
bound.
This behavior is undesirable for two reasons.
First, GP inference scales as $\mathcal{O}(n^3)$, where $n$ is the number of
samples.
Second, old points from outdated regimes can dominate, forcing the surrogate
to confidently explain behavior that is no longer relevant, making
adaptation sluggish.
To eliminate this effect, we maintain a rolling window of samples of size
$W_s$.

\Para{Filtering data.}
As \ours converges to a good solution, every new sample is likely to lie in the
same small neighborhood, and thus nearly identical in the input space.
Updating a GP requires inverting a covariance matrix, which can become
ill-conditioned if most of its rows are almost the same.
This can cause the GP posterior mean and variance to become unreliable.
\ours employs a simple data admission policy to mitigate this risk.
We say a configuration is \emph{novel} if it is far enough away from all other
configurations in the rolling window of samples, and we say a sample is
\emph{surprising} if the GP mispredicts its outcome.
\ours admits a sample if at least one of these conditions is met.
This rule acts as a lightweight guard that keeps the GP numerically stable.

\Para{Denoising data.}
Real systems are noisy, especially when measuring metrics like tail latency.
Feeding jittery data into the GP can lead it to interpret noise as signal,
slowing learning, and chasing spurious spikes.
Before updating the GP with a new sample, \ours applies a causal low-pass
Gaussian smoother whose width adapts to the volatility of the signal.
If the signal becomes completely stable, the filter automatically turns off.
We evaluate the effect of the denoiser in \Cref{s:eval-ablation}.

\rev{
\subsection{Measuring workloads online} \label{s:workfeed}

Recall that the performance models \ours uses require workload
estimates---representative sequences of flows with their sizes, classes, and
arrival times---as input.
In a live network, obtaining these estimates requires capturing flow-level metrics from production traffic.
We built \texttt{Workfeed}, a lightweight distributed system that monitors
ongoing traffic and provides workload traces for model evaluation.
\texttt{Workfeed} measures TCP flows to mirror the workload generator; extending it to capture RPC-level traces over long-lived connections is future work.

\Para{Flow-level monitoring.}
\texttt{Workfeed} uses eBPF probes attached to TCP socket lifecycle events (\texttt{inet\_sock\_set\_state}, \texttt{tcp\_destroy\_sock}) to capture flow metadata.
For each completed flow, the system records the 5-tuple, DSCP marking (for
traffic class discrimination), bytes transferred, and timestamps.
The probes operate with minimal overhead and use a zero-copy ring buffer design for efficiency.

\Para{Rack-level aggregation.}
Flow records are collected from individual hosts and aggregated at the rack level.
Each rack runs a aggregator process that receives flow records via UDP and
forwards them to the controller.
Optionally, the aggregator can downsample flows using deterministic hash-based
sampling and assign weights for statistical reconstruction, though in our
evaluation, we capture complete flow traces and replay them exactly.

\Para{Performance impact.}
Workfeed's kernel-level monitoring introduces negligible overhead.
Each flow record is compact (48 bytes), and the system is designed to support
tracking tens of thousands of concurrent flows per host.
In our dynamic adaptation experiments, \texttt{Workfeed} successfully captured
workloads without measurable impact on application latency or throughput.

}
\section{Limitations}
\label{s:limitations}

This section discusses \ours's limitations.


\paragraph{No formal guarantees.}
\ours provides no formal guarantees of convergence to globally optimal configurations.
Convergence depends on model quality and the effectiveness of the sampling strategy.
The trust region ensures monotone improvement in expectation inside one regime,
but global optimality is not guaranteed.
This safety-performance tradeoff is typical of local optimization methods, which
prioritize safe, incremental improvements over global exploration.

\paragraph{Scaling with dimensionality.}
As the number of adjusted knobs increases, so does the sample complexity
required for accurate modeling and optimization.
While \ours uses Gaussian Process mixtures~\cite{Lafage2022} to scale to
moderate dimensions, they remain inherently limited by the curse of
dimensionality.
This limits \ours's scalability in very high-dimensional control problems without
additional techniques like dimensionality reduction or structured modeling.

\paragraph{No categorical knobs.}
\ours currently only handles continuous parameters; categorical choices are not
supported.
A straightforward extension is to one-hot encode each category and treat the
resulting binary dimensions as additional continuous inputs.
We leave this to future work.

\paragraph{Controller hyperparameters.}
As currently formulated, \ours introduces a set of hyperparameters,
including data window size, trust region size, regime shift
detector thresholds, and others.
All experiments in this paper use a single, fixed setting for each.
In practice, we have found that most parameters could be set once and did not
require tuning.
Some, like the trust region radius $\epsilon$, data window size $W_s$, and regime
shift thresholds $\phi$ and $k$ may need to be adjusted depending on model
quality and traffic volatility.
%
\section{Implementation}
\label{s:implementation}

We implement \ours as a modular controller framework in Rust encompassing workload generation, performance modeling, runtime correction, optimization, and system actuation.

\paragraph{Workload generation.}
We develop \texttt{emu}, a workload generator that emits TCP flows
using configurable workloads (e.g., flow size and inter-arrival distributions).
Traffic classes are marked with DSCP values (10, 18, 26), enabling per-class policy control.
\texttt{emu} uses Prometheus~\cite{prometheus} to collect aggregate flow metrics
(e.g., p99 FCT slowdown) to feed to the controller.

\paragraph{Performance models.}
We extend both \pmn~\cite{zhao2023scalable} and \mthree~\cite{li2024m3} to
operate over a 9-dimensional parameter space consisting of switch weights,
initial congestion windows, and DCTCP marking thresholds for three
traffic classes.
\mthree is a machine learning-based predictor trained on data collected from
the target deployment environment (\cloudlab or \nsthree, depending on the
experiment) using Facebook's public traces~\cite{fb-network}.
For Gaussian processes in the corrective overlay, we use the
\texttt{egobox-moe}~\cite{Lafage2022} library, automatically selecting the
best-fit kernel and mean function using the Bayesian Information Criterion.

\paragraph{System actuation.}
For experimentation in ~\cloudlab~\cite{duplyakin2019design}, we implement
actuators for per-class switch weights, DCTCP marking thresholds, and initial
congestion windows.
The first two are configured via the DellS4048 switch's CLI; the last uses the \texttt{ip route} utility.
We also add support for tuning these parameters in \nsthree~\cite{ns-3} by
adapting the HPCC codebase~\cite{hpcc-codebase}.

\section{Evaluation} \label{s:eval}

To evaluate \ours's (\texttt{pol}) performance, we ask:

\begin{itemize}[leftmargin=*, noitemsep, topsep=0pt]
  \item Does \ours converge to meet SLOs?
  \item How long does convergence take?
  \item Can \ours adapt to changing conditions?
    %
  \item How does \ours's performance depend on the speed and accuracy of its performance model?
  \item Which of \ours's components influences its performance the most?
\end{itemize}

%
We use \cloudlab~\cite{duplyakin2019design} to evaluate \ours on real machines, where performance depends on real transport protocols, NIC hardware, and system-level scheduling noise.
This setting exposes the controller to a range of variability present in real deployments, capturing effects that models and simulations omit.

While \cloudlab provides a realistic testbed, it is also resource-constrained, especially in the number of bare-metal switches.
To test larger topologies with more diverse workloads, we perform a scalability analysis with \nsthree~\cite{ns-3} as the ground truth.

\subsection{Setup}
\begin{table}
  \footnotesize
  \centering
  \begin{tabular}{llll}
    \toprule
    Variant   & Description & Model acc. & $\Delta t$ \\
    \midrule
    \polyp   & Polyphony/\pmn~\cite{zhao2023scalable}         & Low    & $\sim$95s    \\
    \polym    & Polyphony/\mthree~\cite{li2024m3}     & High   & $\sim$7s \\
    \polys       & \polys baseline~\cite{somashekar2024oppertune,karthikeyan2023selftune}     & ---    & $<$1ms  \\
    \bottomrule
  \end{tabular}
  \caption{\small
    Controller variants used in the evaluation.
    $\Delta t$ is the amount of time it takes to compute the next
    configuration after receiving feedback from \cloudlab.
  }
  \vspace{-8mm}
  \label{tab:controller-variants}
\end{table}

\Para{Classes and knobs.}
Across experiments, we split traffic into three classes, corresponding to Differentiated Services Code Point (DSCP) values 10, 18, and 26.
Each class is assigned its own configuration of three control knobs: switch weight, DCTCP marking threshold
K, and initial congestion window (CWND).
These knobs span different layers of the stack  and together define a 9-dimensional configuration space.

\Para{Variants and baseline.}
\Cref{tab:controller-variants} shows two variants of \ours using different performance models and the baseline.
\polyp uses \pmn~\cite{zhao2023scalable} as the performance model.
This shows how the controller behaves when the performance model is far slower
and less accurate.
In \Cref{fig:model-error}, we saw that \pmn had 62\% average error on
\cloudlab, and in our controller experiments, we observe that each controller
iteration takes a minute and a half.
Compared to \polyp, \polym uses a much more accurate machine learned model
\mthree, which saw 17.6\% average error in \Cref{fig:model-error} and reduces
the controller iteration time to seven seconds.
Lastly, we configure \polys as a baseline.

\Para{Configuring \polys.}
\polys has hyperparameters $\delta$ and $\eta$, which are the perturbation radius and learning rate, respectively.
The authors recommend setting $\delta = {O}(1)$ and $\eta = {O}(\delta^2)$, so we choose $\delta = 0.1$ and $\eta = 0.01$.
To specify a reward function, we reuse our objective from \Cref{s:methods-def}, but we rescale it to be in $(-1, 1)$ using the hyperbolic tangent function.
If $g$ is the value of the objective we're trying to minimize, we set the reward to $\tanh(-\beta g)$.
The parameter $\beta$ is tuned to typical values of the objective to prevent the reward from saturating too early or responding too weakly.
In our experiments, we set $\beta = 0.3$.
To make \polys tune simplex parameters, we replicate \ours's strategy: we tune the parameters in unconstrained space and then apply softmax to map them back to the simplex.
Finally, we have found \polys to be sensitive to noisy measurements.
To improve its stability, we wait longer for the objective to settle before computing the reward -- two minutes vs. the default one minute.

\Para{Metrics.}
We evaluate each controller variant using four key metrics.
First, we track the optimization objective over time, which is a unified metric for SLO compliance and fairness.
Second, we report \emph{convergence time}, defined as the first point after which the system meets all SLOs for $k=3$ consecutive iterations.
Third, we measure \emph{hindsight regret}, which integrates the difference
between the actual objective and the best achievable in hindsight,
penalizing exploration that does not improve the objective.
%
Lastly, we compute \emph{min-max fairness} at the end of each experiment, defined as the ratio of the smallest to largest normalized slowdown across classes.

\Para{\cloudlab setup.}
We run our experiments on five \texttt{xl170} machines connected via 10G links to a Dell S4048-ON switch.
One machine acts as the manager, while the remaining four serve as traffic endpoints: one receiver and three senders.
Each machine runs \texttt{emu} (\Cref{s:implementation}), our workload generator, which issues TCP flows between senders and receiver according to specified traffic patterns.
This environment includes real NICs, full transport stacks, and operating system-level variability, allowing us to evaluate the controller under realistic sources of noise and nondeterminism.


\subsection{Convergence study}\label{s:eval-convergence}
\begin{figure*}[h]
  \centering
  \begin{subfigure}[b]{0.25\textwidth}
    \includegraphics[width=\textwidth]{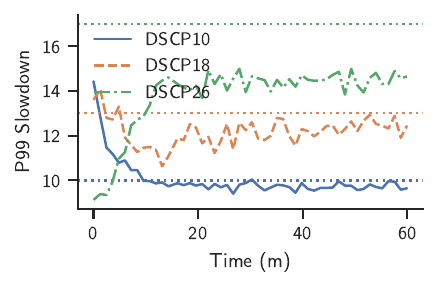}
    \caption{\polym}
    \label{fig:convergence-slowdown-1}
  \end{subfigure}
  \begin{subfigure}[b]{0.25\textwidth}
    \includegraphics[width=\textwidth]{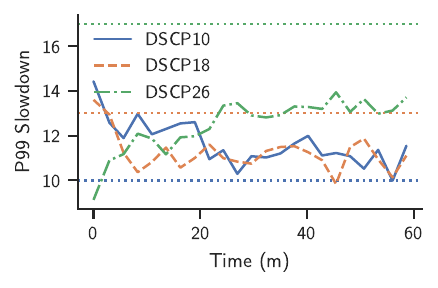}
    \caption{\polyp}
    \label{fig:convergence-slowdown-2}
  \end{subfigure}
  \begin{subfigure}[b]{0.25\textwidth}
    \includegraphics[width=\textwidth]{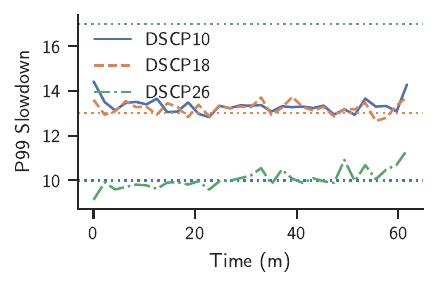}
    \caption{\polys}
    \label{fig:convergence-slowdown-3}
  \end{subfigure}
  \vspace{-2mm}
  \caption{\small Per-class 99th percentile flow completion time slowdowns under high SLO constraints.
    \polym meets all SLOs within a few iterations. \polyp slowly trends toward
    SLO conformance but does not meet the SLO for DSCP 10. Finally, \polys
    misses SLOs for two of the three classes, showing little improvement when
    relying on noisy gradient estimates.
  }
  \label{fig:convergence-slowdown}
\end{figure*}

\begin{figure*}[t]
  \centering
  \begin{minipage}[t]{0.25\textwidth}
    \centering
    \includegraphics[width=\linewidth]{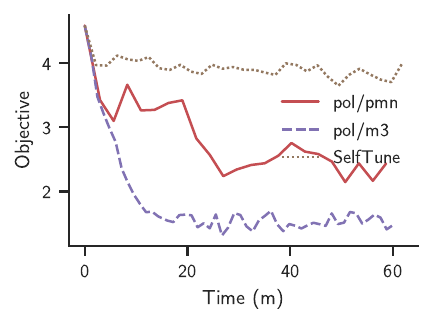}
    \caption{\small Objective over time under high SLO constraints.}
    \label{fig:convergence-objective}
  \end{minipage}
  \hfill
  \begin{minipage}[t]{0.72\textwidth}
    \centering
    \begin{subfigure}[b]{0.32\textwidth}
      \includegraphics[width=\linewidth]{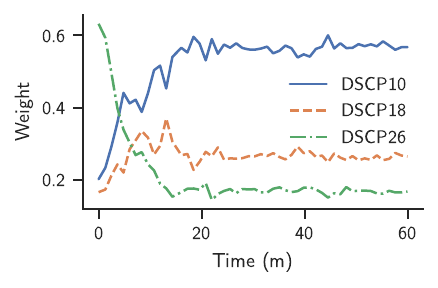}
      \caption{Switch weights}
      \label{fig:convergence-weight}
    \end{subfigure}
    \hfill
    \begin{subfigure}[b]{0.32\textwidth}
      \includegraphics[width=\linewidth]{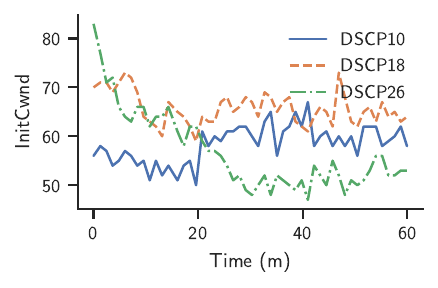}
      \caption{Initial CWND}
      \label{fig:convergence-cwnd}
    \end{subfigure}
    \hfill
    \begin{subfigure}[b]{0.32\textwidth}
      \includegraphics[width=\linewidth]{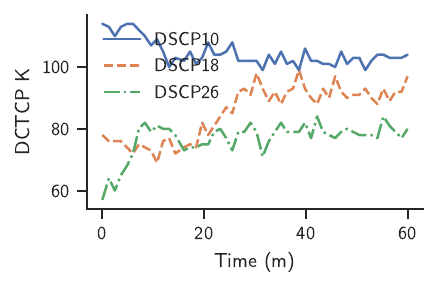}
      \caption{DCTCP K}
      \label{fig:convergence-k}
    \end{subfigure}
    \caption{\small Parameter trajectories for \polym under high SLO
      constraints. \ours coordinates switch and host tuning to meet SLOs and
    fairness objectives.}
    \label{fig:convergence-params}
  \end{minipage}
\end{figure*}

We begin by evaluating 1) whether \ours can discover configurations that
satisfy SLOs and 2) how long convergence takes.
This test evaluates convergence against a fixed, known workload with three
traffic classes and varying levels of SLO stringency.
In \Cref{s:eval-adaption}, we evaluate \ours's ability to adapt to dynamically changing workloads from \texttt{Workfeed}.

We use publicly released workloads from Meta's data center network~\cite{fb-network}.
The flow size distributions are sampled from applications like databases, web servers, and Hadoop.
Inter-arrival times follow a log-normal distribution with a $\sigma$ = 2 for bursty traffic~\cite{zhao2023scalable}.
Flows are evenly distributed across DSCP classes 10, 18, and 26, each contributing 20\% for a network load 60\%.
Experiments run for 60 minutes under three levels of SLO tightness, specified as thresholds for 99th percentile slowdown---one per DSCP:
\textit{Low}: (12, 16, 22), \textit{Medium}: (11, 15, 20), \textit{High}: (10, 13, 17).

\Cref{tab:convergence-multi} summarizes convergence time and objectives across low, medium, and high SLO tightness levels.
\Cref{fig:convergence-slowdown} shows the 99th percentile flow completion time slowdowns for each traffic class over time, under high SLO tightness.
\polym (\Cref{fig:convergence-slowdown-1}) converges within 9.5 minutes, satisfying all per-class SLOs.
Its behavior remains stable and consistent across traffic classes.
In contrast, \polyp (\Cref{fig:convergence-slowdown-2}) converges slower due to
a slower and less accurate model.
While it stabilizes for DSCP 18 and 26, it fails to meet the SLO for DSCP 10.
In \Cref{fig:convergence-slowdown-3}, \polys struggles to meet the SLOs for
DSCP 10 and 18, oscillating without converging to better settings over the
course of the run.
This illustrates the challenge of optimizing under noise and high dimensionality without
a performance model to guide search.

\begin{table}[t]
\centering
\footnotesize
\begin{tabular}{lcrrr}
\toprule
\textbf{Variant} & \textbf{Converge Time} & \textbf{Regret} & \textbf{Fairness} & \textbf{Final Obj.} \\
\midrule
\multicolumn{5}{l}{\textbf{Low SLO tightness}} \\
\polym & 3.7 mins & 21.0 & 0.83 & 1.49 \\
\polyp & 38.2 mins & 67.7 & 0.66 & 2.01 \\
\polys & Not converged & 89.7 & 0.57 & 3.30 \\
\midrule
\multicolumn{5}{l}{\textbf{Medium SLO tightness}} \\
\polym & 6.0 mins & 20.7 & 0.86 & 1.43 \\
\polyp & Not converged & 83.2 & 0.72 & 2.09 \\
\polys & Not converged & 127.7 & 0.46 & 3.15 \\
\midrule
\multicolumn{5}{l}{\textbf{High SLO tightness}} \\
\polym & 9.5 mins & 24.7 & 0.89 & 1.47 \\
\polyp & Not converged & 79.9 & 0.70 & 2.44 \\
\polys & Not converged & 154.4 & 0.47 & 3.99 \\
\bottomrule
\end{tabular}
\caption{\small Convergence metrics across SLO tightness levels in \cloudlab.}
\vspace{-8mm}
\label{tab:convergence-multi}
\end{table}

\Cref{fig:convergence-objective} presents the global objective evolution under
high SLO tightness for three controller variants.
\polym improves quickly and stabilizes in 10 minutes
\polyp achieves gradual objective reduction, but its trajectory flattens early
and does not converge within the 60-minute window due to the slower and less
accurate model.
\polys improves the objective slightly but struggles to extract a meaningful
gradient signal from noisy measurements.
%

\begin{figure*}[h]
\centering
\begin{subfigure}[b]{0.25\textwidth}
\includegraphics[width=\textwidth]{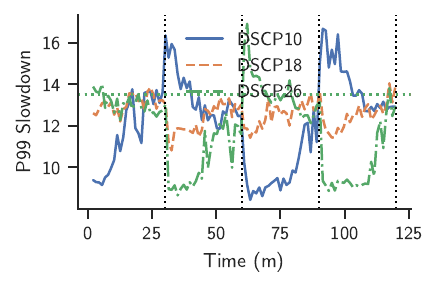}
\caption{\polym}
\label{fig:adapt-slowdown-1}
\end{subfigure}
\begin{subfigure}[b]{0.25\textwidth}
\includegraphics[width=\textwidth]{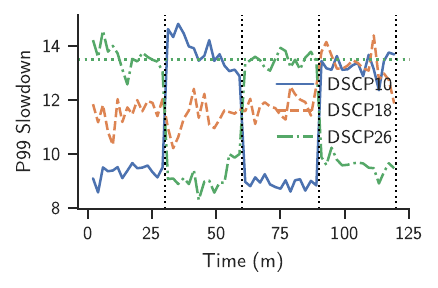}
\caption{\polyp}
\label{fig:adapt-slowdown-2}
\end{subfigure}
\begin{subfigure}[b]{0.25\textwidth}
\includegraphics[width=\textwidth]{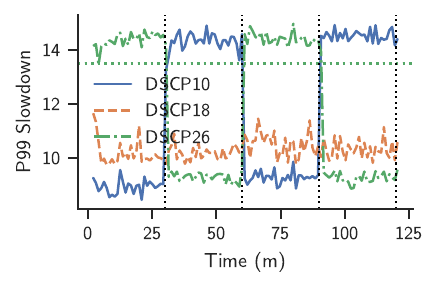}
\caption{\polys}
\label{fig:adapt-slowdown-3}
\end{subfigure}
\vspace{-2mm}
\caption{
\small Per-class 99th percentile flow completion time slowdowns under workload shifts.
\polym adapts within minutes to reestablish SLO compliance.
\polyp converges slowly and has a higher proportion of missed SLOs.
\polys does not meet SLOs and shows no clear adaptation.
}
\label{fig:adapt-slowdown}
\end{figure*}

\begin{figure*}[t]
\centering
\begin{minipage}[t]{0.25\textwidth}
\centering
\includegraphics[width=\linewidth]{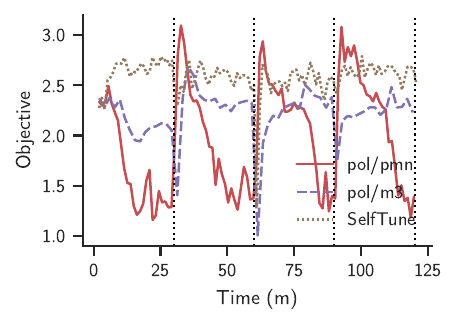}
\caption{\small Objective over time under regime shifts.}
\label{fig:adapt-objective}
\end{minipage}
\hfill
\begin{minipage}[t]{0.72\textwidth}
\centering
\begin{subfigure}[b]{0.32\textwidth}
  \includegraphics[width=\linewidth]{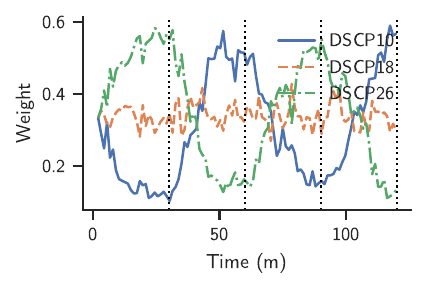}
  \caption{Switch weights}
  \label{fig:adapt-weight}
\end{subfigure}
\hfill
\begin{subfigure}[b]{0.32\textwidth}
  \includegraphics[width=\linewidth]{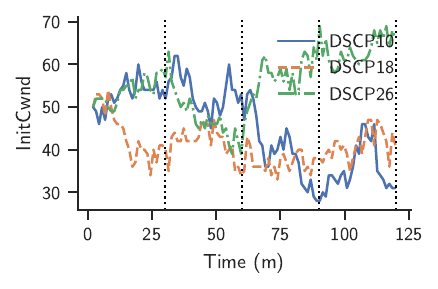}
  \caption{Initial CWND}
  \label{fig:adapt-cwnd}
\end{subfigure}
\hfill
\begin{subfigure}[b]{0.32\textwidth}
  \includegraphics[width=\linewidth]{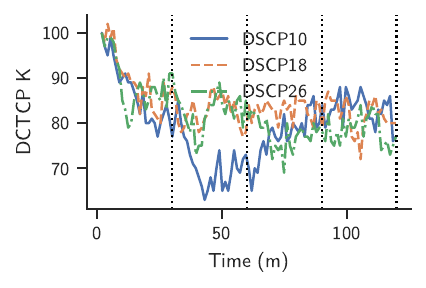}
  \caption{DCTCP K}
  \label{fig:adapt-k}
\end{subfigure}
\caption{\small Parameter evolution under \polym. \ours gradually updates parameters in response to workload changes.}
\label{fig:adapt-params}
\end{minipage}
\end{figure*}

\Cref{fig:convergence-params} shows the evolution of parameters under \polym: switch queue weights, initial CWNDs, and DCTCP marking thresholds.
All parameters gradually adjust and remain stable over time.
We observe that \ours jointly optimizes host and network settings to balance
fairness and performance. Note that the optimal initial window size depends on
the scheduling weight, so that the medium traffic class gets the largest window
size. A large initial window with high weight will negatively impact the SLOs
for other traffic classes.

%

\subsection{Adaptation study}\label{s:eval-adaption}

This section evaluates whether \ours can adapt its control decisions in the presence of workload changes measured live from \texttt{Workfeed}.

In this experiment, we use the same bursty traffic workloads as described
in~\Cref{s:eval-convergence}, and we impose a uniform SLO of 13.5 on all
classes.
To test adaptability, we introduce three regime shifts during a two hour
experiment, and we use \texttt{Workfeed} to measure the workload and send it to
the controller.
\Cref{tab:adaptation-loads} summarizes the offered load profile across four 30-minute phases.
Each phase requires the controller to reconfigure both host and switch-level parameters to maintain per-class SLO compliance.

\begin{table}[t]
\centering
\footnotesize
\begin{tabular}{lrrr}
\toprule
\textbf{Phase} & \textbf{DSCP 10} & \textbf{DSCP 18} & \textbf{DSCP 26} \\
\midrule
I: 0--30 min  & 500 & 2000 & 3500 \\
II: 30--60 min  & 3500 & 2000 & 500 \\
III: 60--90 min & 500 & 2000 & 3500 \\
IV: 90--120 min & 3500 & 2000 & 500 \\
\bottomrule
\end{tabular}
\caption{\small Traffic load profiles used in the adaptation study. Offered load (in Mbps) varies every 30 minutes to challenge controller responsiveness.}
\vspace{-8mm}
\label{tab:adaptation-loads}
\end{table}

\Cref{fig:adapt-slowdown} shows the per-class 99th percentile flow slowdowns throughout the adaptation experiment.
\polym (\Cref{fig:adapt-slowdown-1}) adapts quickly to workload shifts, reestablishing SLOs and equalizing classes within minutes.
\polyp (\Cref{fig:adapt-slowdown-2}) adapts slower, taking nearly a half hour in some cases to reach the SLO threshold.
It does not converge fast enough to equalize classes within any half hour window.
\polys (\Cref{fig:adapt-slowdown-3}) does little to improve performance and does not respond to workload changes.

\Cref{fig:adapt-objective} shows the global objective over time.
\polym quickly reduces the objective quickly after each workload shift.
While \polyp steers away from large SLO violations, it scores poorly in the fairness dimension, so the objective is persistently higher.
\polys does not reduce the objective throughout the experiment.

Lastly, \Cref{fig:adapt-params} shows how \polym's parameter evolution over time.
Most interpretable are the switch weights (\Cref{fig:adapt-weight}), which
shift predictably to favor under-served classes in each regime (e.g., DSCP 10
in Phase II).


\begin{figure*}[h]
\centering
\begin{subfigure}[b]{0.24\textwidth}
\includegraphics[width=\textwidth]{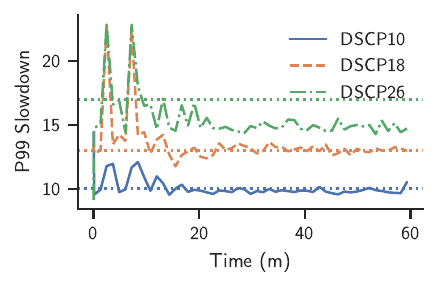}
\caption{No trust region}
\label{fig:ablate-main-1}
\end{subfigure}
\begin{subfigure}[b]{0.24\textwidth}
\includegraphics[width=\textwidth]{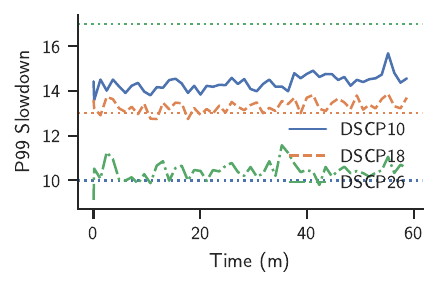}
\caption{No correction}
\label{fig:ablate-main-2}
\end{subfigure}
\begin{subfigure}[b]{0.24\textwidth}
\includegraphics[width=\textwidth]{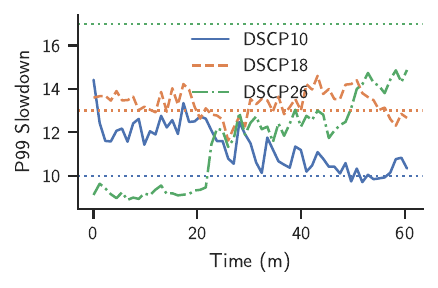}
\caption{No model}
\label{fig:ablate-main-3}
\end{subfigure}
\begin{subfigure}[b]{0.24\textwidth}
\includegraphics[width=\textwidth]{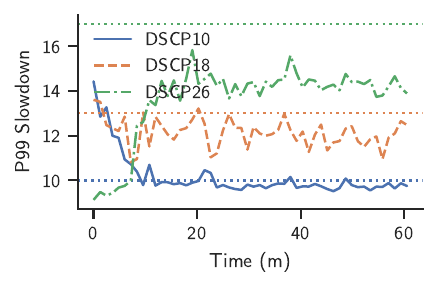}
\caption{No denoiser}
\label{fig:ablate-main-4}
\end{subfigure}

\begin{subfigure}[b]{0.24\textwidth}
\includegraphics[width=\textwidth]{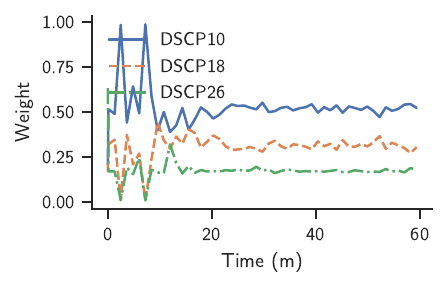}
\caption{No trust region}
\label{fig:ablate-main-6}
\end{subfigure}
\begin{subfigure}[b]{0.24\textwidth}
\includegraphics[width=\textwidth]{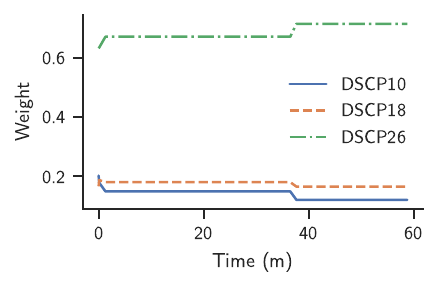}
\caption{No correction}
\label{fig:ablate-main-7}
\end{subfigure}
\begin{subfigure}[b]{0.24\textwidth}
\includegraphics[width=\textwidth]{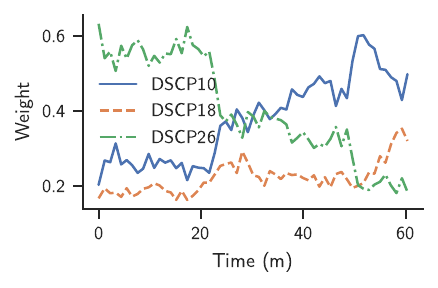}
\caption{No model}
\label{fig:ablate-main-8}
\end{subfigure}
\begin{subfigure}[b]{0.24\textwidth}
\includegraphics[width=\textwidth]{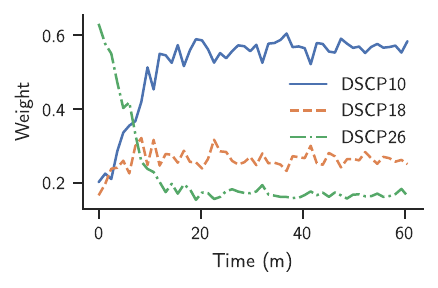}
\caption{No denoiser}
\label{fig:ablate-main-9}
\end{subfigure}
\vspace{-2mm}
\caption{\small Per-class 99th percentile slowdowns (top) and switch weights (bottom) under ablation.
Removing key components—like the trust region, performance model, or correction—degrades SLO compliance and stability.
Note: Y-axis scales vary.}
\label{fig:ablate-main}
\end{figure*}

\begin{figure}[h]
\centering
\includegraphics[width=0.35\textwidth]{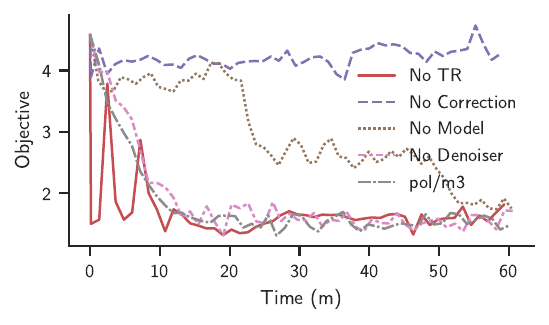}
\vspace{-3mm}
\caption{\small Global objective over time under component ablations. The full
\ours controller (\texttt{pol/m3}) and no denoiser both converge; no trust
region converges but with oscillation. No model converges more slowly, and no
correction does not make progress.}
\label{fig:ablate-objectives}
\vspace{-8mm}
\end{figure}

\subsection{Ablation study}
\label{s:eval-ablation}

We use the same bursty workload and high tightness SLO thresholds as in~\Cref{s:eval-convergence}.
All experiments are based on \polym, but each omits a specific component:

\begin{itemize}[leftmargin=*, noitemsep]
\item \textit{No trust region (TR)} disables the trust region.
\item \textit{No correction} disables the online model correction, relying solely on \mthree's predictions.
\item \textit{No model} removes the model (\mthree), forcing the controller to explore directly on the real system.
\item \textit{No denoiser} feeds raw (unsmoothed) slowdown measurements to the controller.
\end{itemize}

\Cref{fig:ablate-main} presents per-class slowdown trajectories and corresponding weight adjustments.
%
%
Removing the trust region (\Cref{fig:ablate-main-1}) results in larger
parameter changes and early SLO violations (especially for DSCP 18 and 26), with
weight oscillations between extremes.
Without residual correction (\Cref{fig:ablate-main-2}) the biased simulator mis-orders nearby points,
and fails to improve the objective within a trust region.
Without the model (\Cref{fig:ablate-main-3}), the controller must explore many more configurations before it finds a good one, missing SLOs for the entire duration.
The ``no denoiser" variant performs well but exhibits high-frequency jitter in slowdowns and weights.

\Cref{fig:ablate-objectives} summarizes the global objective under each variant.
\polym converges smoothly.
"No model" and "no correction" both plateau early, with the latter performing the worst overall.
"No TR" reduces the objective, but with high initial variance.

%

\subsection{Scalability analysis in \nsthree}
\label{s:eval-scalability}

\begin{figure}[t]
\centering
\includegraphics[width=0.6\linewidth]{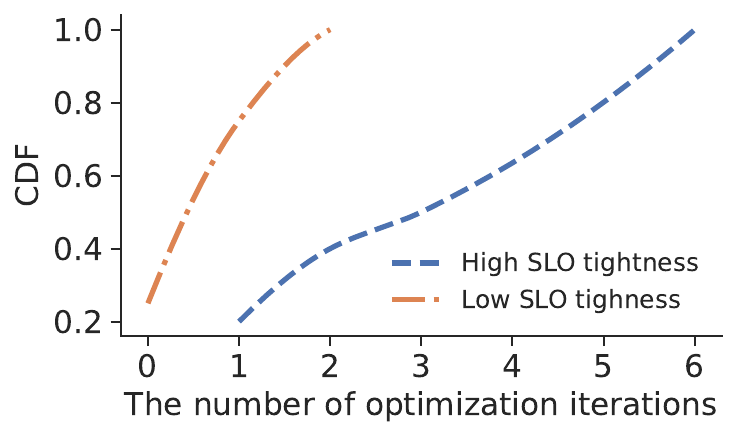}
\vspace{-3mm}
\caption{\small CDF of convergence time across \nsthree scenarios; scenarios that do not converge are omitted.}
\label{fig:sim-cdf}
\vspace{-4mm}
\end{figure}

\begin{figure*}[h]
\centering
\begin{subfigure}[b]{0.19\textwidth}
\includegraphics[width=\textwidth]{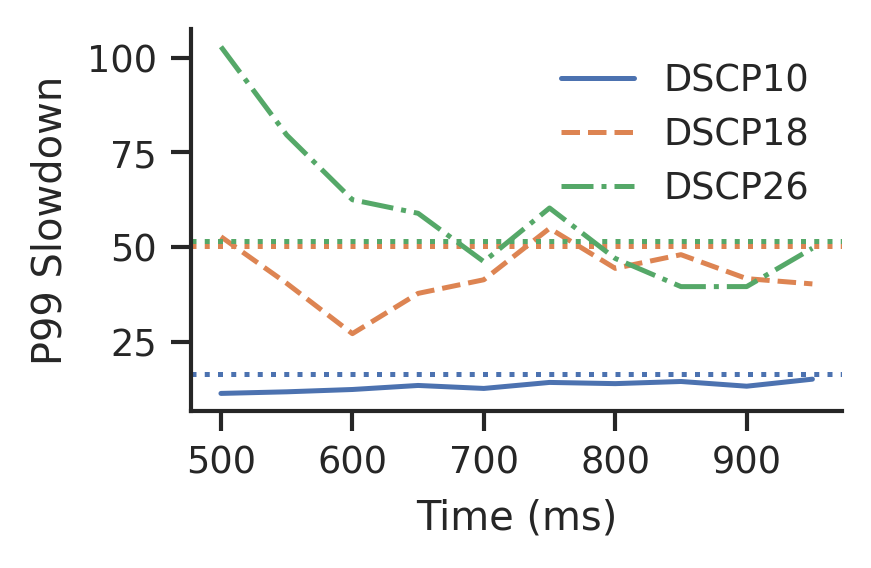}
\caption{Slowdown (p99)}
\label{fig:sim-1}
\end{subfigure}
\begin{subfigure}[b]{0.19\textwidth}
\includegraphics[width=\textwidth]{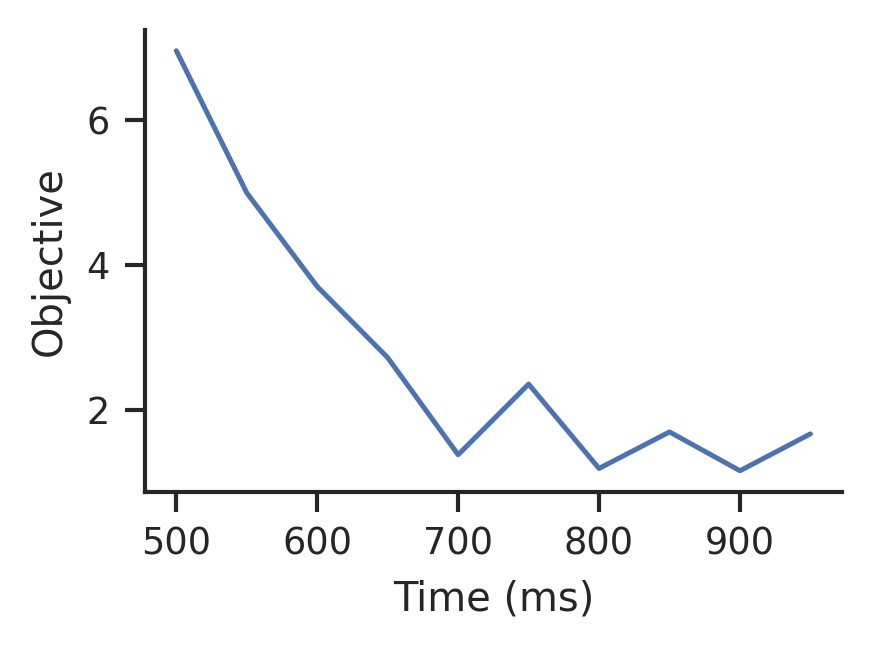}
\caption{Objective value}
\label{fig:sim-2}
\end{subfigure}
\begin{subfigure}[b]{0.19\textwidth}
\includegraphics[width=\textwidth]{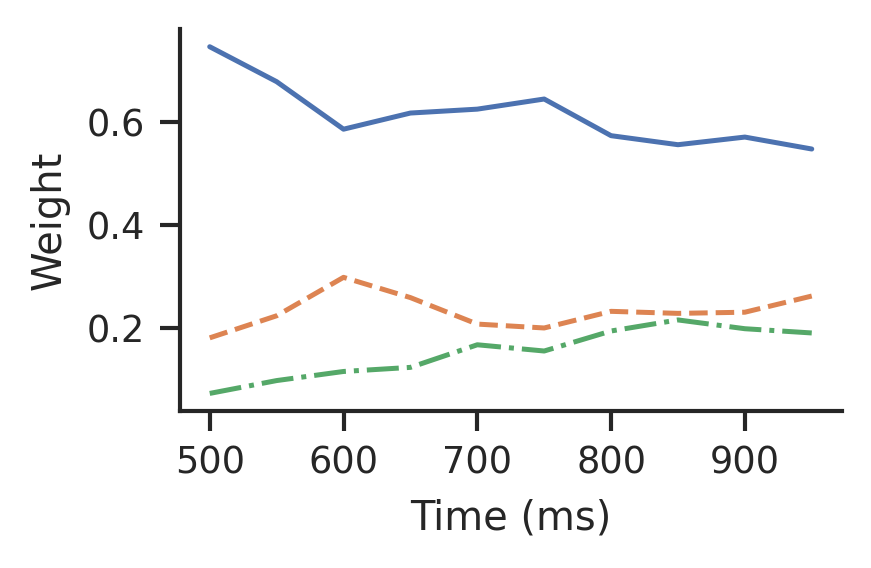}
\caption{Switch weights}
\label{fig:sim-3}
\end{subfigure}
\begin{subfigure}[b]{0.19\textwidth}
\includegraphics[width=\textwidth]{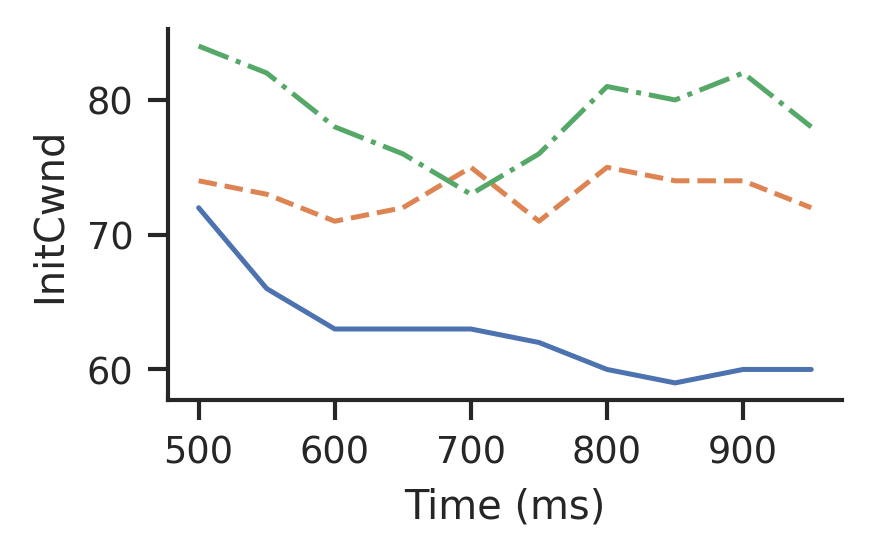}
\caption{Initial CWND}
\label{fig:sim-4}
\end{subfigure}
\begin{subfigure}[b]{0.19\textwidth}
\includegraphics[width=\textwidth]{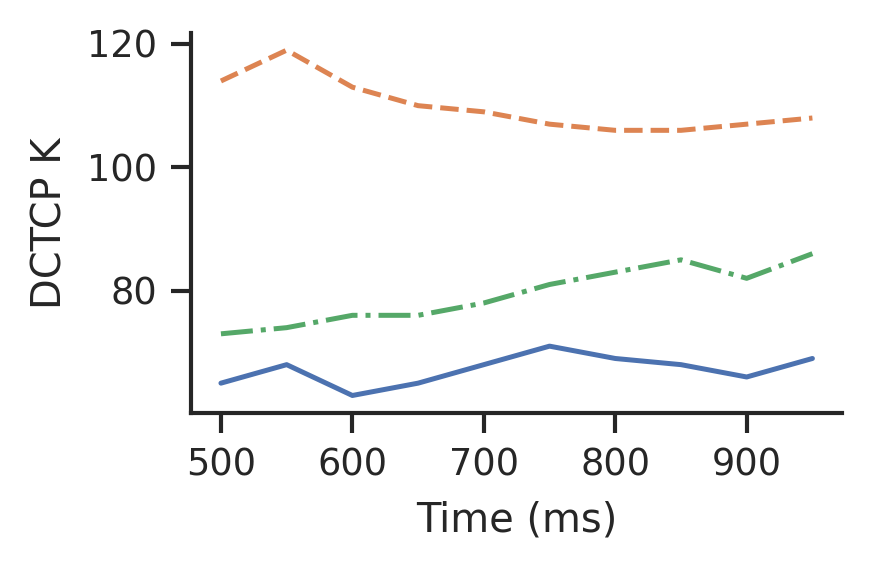}
\caption{DCTCP K}
\label{fig:sim-5}
\end{subfigure}
\vspace{-2mm}
\caption{
\small \polym's behavior in \nsthree~\cite{ns-3}.
\textbf{(a)} Per-class 99th percentile flow completion time slowdowns; dotted lines indicate SLO thresholds.
\textbf{(b)} Global objective improves steadily over time.
\textbf{(c–e)} Controller updates switch weights, initial congestion windows, and DCTCP thresholds smoothly to meet SLOs.
}
\label{fig:sim}
\end{figure*}

While our \cloudlab~\cite{duplyakin2019design} experiments validate \ours in realistic and noisy conditions, they are limited to a small topology.
We conduct a scalability analysis using \nsthree~\cite{ns-3}.

We construct 24 simulation scenarios by varying the following parameters:
\begin{itemize}[leftmargin=*, noitemsep]
\item \textit{Topology}: We use a 32-rack, 256-host fat-tree topology sampled from Meta's data center networks~\cite{fb-fabric}.
\item \textit{Workload:} We generate flow-level traffic using two flow size distributions derived from publicly available production workloads (\texttt{WebSearch} and \texttt{Hadoop}~\cite{fb-network}).
Inter-arrival times follow a log-normal distribution with $\sigma = 1$ (low burstiness) or $\sigma = 2$ (high burstiness)~\cite{zhao2023scalable}.
The network load is fixed at 60\%.
We use three application-specific traffic matrices, database, web server, and Hadoop clusters~\cite{fb-network}, to capture realistic rack-to-rack communication patterns.
Flow endpoints are selected uniformly at random from hosts within each rack.
\item \textit{SLO tightness}: For each scenario, we randomly select an SLO tightness level by scaling a baseline slowdown by 25\% (high tightness) or 50\% (low).
Baseline slowdowns are obtained by running a simulation under a fixed, near-optimal configuration.
The 99th percentile slowdown from this simulation is used as the baseline SLO and scaled accordingly.
\end{itemize}

We evaluate \polym based on two metrics: \emph{SLO compliance rate} and \emph{convergence time} across all scenarios.
Each simulation runs for 1 second. The first 500 ms are used to gather an initial set of datapoints.
Starting at 500 ms, \polym selects a new configuration every 50 ms.

\Cref{fig:sim} visualizes a representative scenario: the controller reduces
99th percentile flow slowdowns (a) and global objective (b) while making
coordinated updates to switch weights (c), initial CWNDs (d),
and DCTCP thresholds (e).
\Cref{fig:sim-cdf} shows the CDF of convergence times across the 19 scenarios where \polym successfully converged.
\polym achieves 80\% (19/24) SLO compliance.
Despite variations in workload characteristics and SLO tightness, \polym stabilizes to SLOs in most cases:
under low SLO tightness, almost all scenarios converge within two optimization iterations;
under high SLO tightness, convergence takes longer but still completes within six iterations in all cases.
\section{Related Work}


\subsection{Tuners and controllers}

\Para{Blackbox autotuners.}
A large body of work automates configuration tuning by treating the system as a
black box and running controlled trials to search high-dimensional knob spaces.
Representative systems include CherryPick~\cite{alipourfard2017cherrypick} for
cloud instance choices, Metis~\cite{li2018metis} and
OtterTune~\cite{van2017automatic} for service/DBMS parameters,
OPPerTune~\cite{somashekar2024oppertune} and
SelfTune~\cite{karthikeyan2023selftune} for post-deployment service and
cluster-manager knobs, AutoSched~\cite{gao2024autosched} for deep-learning
schedulers, and FLASH~\cite{qiu2024flash} as a sample-efficient Bayesian
optimization method.
These approaches are supervisory and episodic: they propose a configuration,
measure the live system, and iterate to convergence---often over minutes to
hours---and can induce temporary regressions during exploration.
They are effective for finding good static settings but are not designed to
regulate a performance metric on a tight feedback loop.

\Para{Online controllers.}
A separate line uses continuous feedback to control a system variable directly.
For example, Autothrottle employs a bi-level design with a contextual-bandit
making minute-by-minute CPU-throttle targets for microservices to satisfy
latency SLOs~\cite{wang2024autothrottle}, and ACC applies deep RL to adjust ECN
thresholds at switches to keep queues short while sustaining
throughput~\cite{yan2021acc}.
These are online regulators of specific mechanisms (CPU throttling; ECN
marking) rather than broad configuration search.
Polyphony sits in this controller category: it targets network QoS metrics
directly (e.g., class-specific p99) and selects actions on a minute timescale,
but, unlike prior online controllers, does so model-guided, using fast
approximate performance models to predict counterfactual outcomes and reduce
the need for risky live experimentation.

\subsection{Simulation-based optimization and control}
Simulation offers a scalable and low-risk means to explore system behavior, especially in complex networking environments.
Modern network simulators use parallelization~\cite{zhao2023scalable,gao2023dons,bai2024unison} and ML techniques~\cite{mimicnet, deepqueuenet,li2024m3,ferriol2023routenet} to model packet-level behavior with improved speed and fidelity.
Despite these advances, a gap remains between simulated and real-world performance due to unmodeled hardware interactions and environmental variability~\cite{filieri2014automated}.
To bridge this gap, ByteDance's Crescent~\cite{gao2024crescent} emulates production switch software in isolated environments, enabling realistic and low-risk evaluation of network configurations.
Using simulations and emulations as predictive models, Model Predictive Control (MPC) offers formal methods for proactive configuration planning.
Zipper~\cite{balasingam2024application}, for instance, applies MPC to dynamically allocate bandwidth in 5G RAN slicing, using model-based forecasts to maintain SLO compliance.
However, conventional MPC approaches require sufficiently accurate predictions;
robust variants exist, but building fast, globally accurate models in data
center networks is an open problem.
\ours addresses this problem with an online corrective overlay to compensate
for modeling errors, and it further applies trust-region-based optimization
strategies to focus exploration in regions where predictions are most reliable.

\section{Conclusion} \label{s:conclusion}
This paper described the design, implementation, and evaluation of \ours{},
a system for controlling tail latency at minute-scale to achieve performance
objectives for different traffic classes---even under dynamic workloads.
\ours{} leverages fast but inaccurate models to estimate the consequence of
applying a proposed configuration change.
To correct for model error, \ours{}
builds and applies a corrective overlay anchored on the best known alternative; Bayesian
optimization is then used within a region around that alternative.
Provided that the prediction model is reasonably accurate, e.g., using machine
learning, \ours{} can quickly converge to a more optimal operating point and
adapt to changing conditions.
%
Under tight SLOs, noisy measurements, and large workload shifts, \ours stabilizes to meet SLOs within fifteen minutes.

\def\bibfont{\normalfont}

\newpage
\bibliographystyle{abbrv}
\bibliography{refs.bib}

\end{document}